\documentclass[lettersize,journal]{IEEEtran}
\usepackage{amsmath,amsfonts}
\usepackage{array}
\usepackage[caption=false,font=normalsize,labelfont=sf,textfont=sf]{subfig}
\usepackage{textcomp}
\usepackage{stfloats}
\usepackage{url}
\usepackage{verbatim}
\usepackage{graphicx}
\usepackage{cite}
\usepackage{amsmath,amsfonts}
\usepackage{algorithmicx}
\usepackage{algpseudocode}
\usepackage{algorithm}
\usepackage{graphicx}
\usepackage{textcomp}
\usepackage{xcolor}
\usepackage{array}
\usepackage{xspace}
\usepackage{enumitem}
\usepackage{pifont}

\newcommand{\circleone}{\ding{182}\xspace}
\newcommand{\circletwo}{\ding{183}\xspace}
\newcommand{\circlethree}{\ding{184}\xspace}
\newcommand{\circlefour}{\ding{185}\xspace}

\newcommand{\etal}{\textit{et al}.\xspace}

\newcommand{\vs}{\textit{v}.\textit{s}.\xspace}
\newcommand{\etc}{\textit{etc}.\xspace}

\newcommand{\xname}{AGON\xspace}

\newcommand{\opname}{nOP\xspace}
\newcommand{\opnames}{nOPs\xspace}

\newcommand{\code}[2]{\textcolor{#1}{\texttt{#2}}}

\algnewcommand{\LeftCommenta}[1]{\Statex \hspace{-1em} \(\triangleright\) #1}
\algnewcommand{\LeftCommentb}[1]{\Statex \hspace{1.3em} \(\triangleright\) #1}

\begin{document}

\title{\xname: Automated Design Framework for Customizing Processors from ISA Documents}

\author{Chongxiao Li,
        Di Huang,
        Pengwei Jin,
        Tianyun Ma,
        Husheng Han,
        Shuyao Cheng,
        Yifan Hao,
        Yongwei Zhao,
        Guanglin Xu,
        Zidong Du,
        Rui Zhang,
        Xiaqing Li,
        Yuanbo Wen,
        Xing Hu,
        Qi Guo
\IEEEcompsocitemizethanks{
\IEEEcompsocthanksitem Chongxiao Li, Pengwei Jin, and Husheng Han are with the State Key Lab (SKL) of Processors, Institute of Computing Technology (ICT), Chinese Academy of Sciences (CAS), Beijing, China, the University of Chinese Academy of Sciences, Beijing, China, and also with Cambricon Technologies.
\IEEEcompsocthanksitem Di Huang, Shuyao Cheng, Yifan Hao, Yongwei Zhao, Guanglin Xu, Rui Zhang, Yuanbo Wen, and Qi Guo are with the SKL of Processors, ICT, CAS, Beijing, China.
\IEEEcompsocthanksitem Tianyun Ma is with the University of Science and Technology of China, Hefei, China, the SKL of Processors, ICT, CAS, Beijing, China, and also with Cambricon Technologies.
\IEEEcompsocthanksitem Zidong Du and Xing Hu are with the SKL of Processors, ICT, CAS, Beijing, China, and Shanghai Innovation Center for Processor Technologies, Shanghai, China.
\IEEEcompsocthanksitem Xiaqing Li is with the Key Laboratory of Big Data and Artificial Intelligence in Transportation (Beijing Jiaotong University), Ministry of Education, Beijing, China.
}
}

\maketitle

\begin{abstract}
Customized processors are attractive solutions for vast domain-specific applications due to their high energy efficiency.
However, designing a processor in traditional flows is time-consuming and expensive. 
To address this, researchers have explored methods including the use of agile development tools like Chisel or SpinalHDL, high-level synthesis (HLS) from programming languages like C or SystemC, and more recently, leveraging large language models (LLMs) to generate hardware description language (HDL) code from natural language descriptions. However, each method has limitations in terms of expressiveness, correctness, and performance, leading to a persistent contradiction between the level of automation and the effectiveness of the design.
Overall, how to automatically design highly efficient and practical processors with minimal human effort remains a challenge.

In this paper, we propose \xname, a novel framework designed to leverage LLMs for the efficient design of out-of-order (OoO) customized processors with minimal human effort.
Central to \xname is the nano-operator function (nOP function) based Intermediate Representation (IR), which bridges high-level descriptions and hardware implementations while decoupling functionality from performance optimization, thereby providing an automatic design framework that is expressive and efficient, has correctness guarantees, and enables PPA (Power, Performance, and Area) optimization.

Experimental results show that superior to previous LLM-assisted automatic design flows, \xname facilitates designing a series of customized OoO processors that achieve on average 2.35 $\times$ speedup compared with BOOM, a general-purpose CPU designed by experts, with minimal design effort. 
\end{abstract}


\section{Introduction}

\IEEEPARstart{C}{ustomized} processors achieve prominent efficiency and are widely adopted in diverse domain-specific applications, such as the Internet of Things (IoT) and multi-media embedded systems~\cite{shahabuddin2021asip,van2006scalable,de2007ultra,liao2012high,liu2005application,eusse2014flexible}.
Though the combination of Central Processing Units (CPUs) and application-specific accelerators offers both high flexibility and efficiency, this approach still relies heavily on manual efforts and incurs high manual costs due to massive demands for diverse customized scenarios.
For example, the construction of a high-performance RISC-V CPU using the widely recognized Chisel language requires approximately 60,000 lines of code~\cite{xu2022towards}, indicating a substantial investment in the design process.

\begin{figure}[t]
    \centering
    \includegraphics[width=1\linewidth]{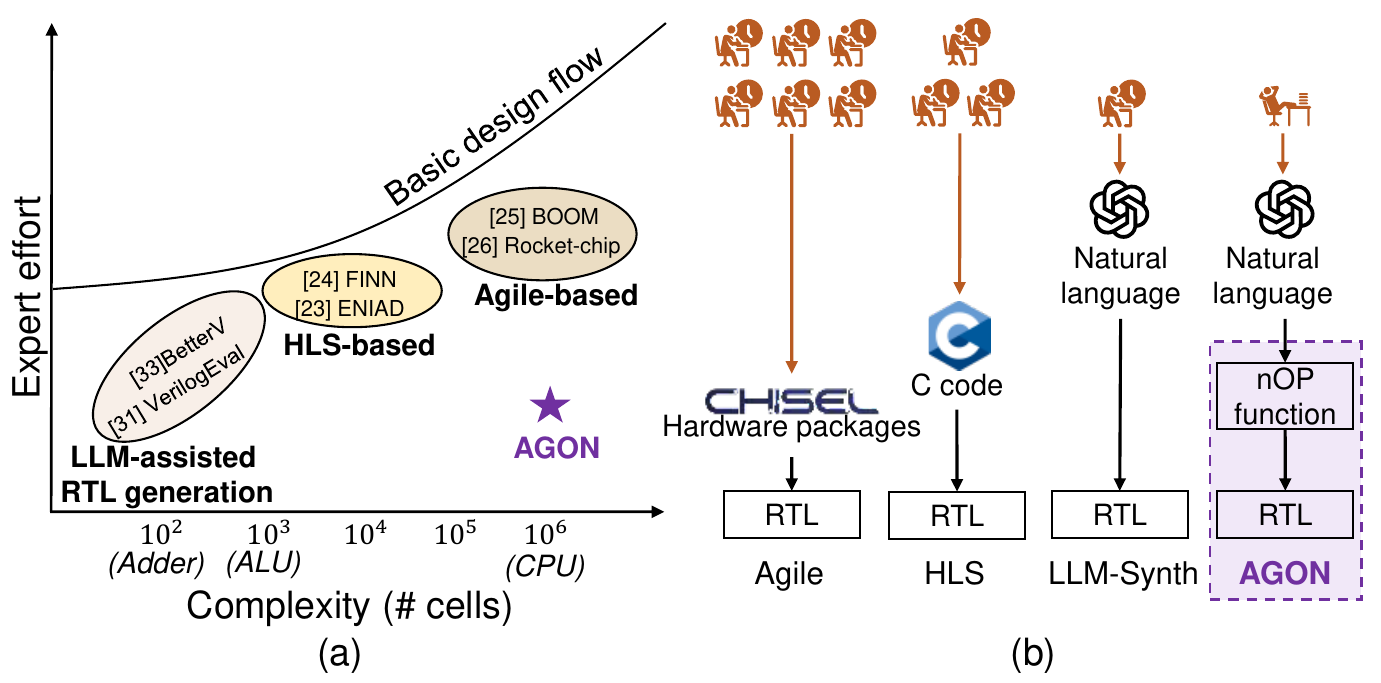}
    \caption{Comparing \xname with existing synthesis work in terms of needed expert effort and (a) complexity, (b) automation level.}
    \label{fig:intro_comparison}
\end{figure}

To reduce the design efforts for customized processors with instruction extensions, researchers and designers have dedicated significant efforts to simplifying or automating the design flow:
\begin{itemize}[labelsep=3pt,leftmargin=10pt]
\item
\textbf{Design on encapsulated hardware abstraction:} Agile development tools simplify the design flow with encapsulated hardware design abstractions. Prominent examples include leveraging Chisel or SpinalHDL in developing hardware systems~\cite{bachrach2012chisel, spinalhdl}. However, it still demands massive expert programming for various customized applications. For instance, constructing a high-performance RISC-V CPU in Chisel necessitates approximately 60k lines of code~\cite{xu2022towards}, representing a significant design effort.  
\item
\textbf{Synthesis from high-level programming language:} High-level synthesis (HLS) tools can automatically synthesize register-transfer level (RTL) hardware codes from high-level languages (C or SystemC) to reduce human effort further. However, the coarse-grained synthesized circuit suffers from sub-optimal performance~\cite{8356004}. Moreover, HLS does not automate the entire process from specifications to circuits. Manually writing high-level language code remains a significant burden.
\item
\textbf{Synthesis from natural language:} Recently, researchers have explored the attempts to utilize large language models (LLMs) to generate RTL code directly from natural language specifications. However, these methods are far from practical use. For example, in RTLLM~\cite{lu2023rtllm}, no LLM-based methods successes in generating a simple 407-cell reduced instruction set computer (RISC) CPU circuits (syntax checking all failed in five trials), including GPT-4. 
This is caused by the following reasons: 
\circleone The gap between LLMs' strength in high-level tasks and processors' low-level hardware implementation.
Hardware Description Languages (HDLs) have a significant {gap} with other high-level programming languages such as Python or C++, due to their explicit control of \emph{port connections}, \emph{sequential structures}, and \emph{signal bit widths}. 
The scarcity of high-quality HDL data (\textasciitilde1GB in HDLs \vs \textasciitilde800GB in all programming languages~\cite{li2023starcoder}) results in LLMs being significantly weaker in HDL generation tasks compared to high-level programming tasks.
Additionally, LLMs struggle with large-scale projects\cite{zhang-etal-2023-repocoder, liu2024repobench, allam2024rtlrepo}, which are exacerbated in processor design due to the extensive code required for managing low-level RTL details.
\circletwo The gap between LLMs' lack of inherent verification capability and the stringent correctness demands in processor design.
Since LLMs generate outputs by predicting the next token based on probability distributions learned from training data, they inherently lack mechanisms to verify the correctness of their generation. However, the hardware design for processors requires high correctness and strict verification. Without efficient verification mechanisms, LLMs cannot generate practical processors correctly, hindering their application in design flows.
\circlethree The gap between functionality description and performance optimization. Beyond accurate design generation, performance optimization is even more challenging for LLMs. Unlike design generation, performance optimization requires a good understanding of functional equivalence and the relationship between HDL code and performance, power, and area (PPA), which is currently unattainable due to the scarcity of training data~\cite{zhong2023llm4eda}.

\end{itemize}

In summary, we are trapped in the contradiction between the level of automation and design effectiveness. As shown in Figure~\ref{fig:intro_comparison}, the higher the level of automation, the simpler, slower, and less efficient the designed circuits. It is challenging for the automation framework to meet all these three requirements: \circleone design complex processors for various applications efficiently, \circletwo achieve functional correctness with minimal human intervention, and \circlethree empower hardware optimization to attain comparable performance with human-crafted processors.

To this end, we introduce \xname, a framework designed for LLM-based design flows, facilitating efficient out-of-order (OoO) customized processor RTL prototyping with minimal human intervention.
\textbf{The key insight} of \xname is a nano-operator function (nOP function) based Intermediate Representation (IR) to not only bridge the gap between high-level descriptions and the hardware-level implementation but also decouple the functionality and the performance optimization. 
\textbf{\xname is not intended to replace existing processor design flows} but to offer an infrastructure for processor development in the era of LLMs. It enables the generation of customized processor RTL from natural language for prototyping, facilitating rapid evaluation and verification, and serves as a basis for further development.
\xname offers the following advantages: 
\circleone \textbf{Being expressive and efficient:} \xname abstracts basic operations in instructions into \opnames, and through the combination of \opnames, it can implement diverse instruction functions. Unlike HDL-based methods, \xname minimizes descriptive code while preserving generality, enhancing LLM generation efficiency. 
\circletwo \textbf{Ensuring correctness:} \xname, through data flow graph representation and bit width inference of \opnames, hides explicit port connection, sequential structures, and bit width management in hardware design from LLMs, enhancing LLM generation accuracy and simplifying manual debugging, thereby minimizing manual intervention. With \opname functions accurately representing instruction functionalities, \xname's primitives employ rule-based functional equivalence transformations to guarantee processor implementation correctness. Additionally, \xname offers multi-level verification primitives to validate the implementation. 
\circlethree \textbf{Enabling PPA optimization:} \xname decouples functionality and optimization, only requiring that LLM generate correct functional descriptions while leveraging multi-level primitives for processor optimization. Among these primitives, \xname also provides PPA-aware auto-tuning, reducing human intervention and achieving optimized design.

The main contributions of this paper are as follows:

\begin{itemize}[labelsep=3pt,leftmargin=18pt]
\item We introduce \xname, an automated framework developed for rapid RTL generation from ISA documents to processors. 
\xname introduces nOP functions IR to assist LLMs efficiently and expressively translating ISA requirements to formal functionality definitions. \xname offers multi-level verification mechanisms for correctness validation.

\item AGON decouples functionality and performance optimization. During transforming nOP functions to RTL design, AGON proposes scheduling primitives and PPA-aware auto-tuning methods for design optimization. AGON offers compiler and simulation tools for automatically generating high-performance processor RTL prototypes.

\item We perform comprehensive evaluations on \xname. The results show that, compared to other LLM-assisted methods, \xname is able to design practical-level processors with minimal human intervention. We showcase designing a series of high-performance out-of-order processors based on AGON. Compared with an expert-designed general-purpose CPU, our processor achieves an average of 2.35 $\times$ speedup on specific tasks.
\end{itemize}

\begin{figure*}[ht]
    \centering
    \includegraphics[width=1\linewidth]{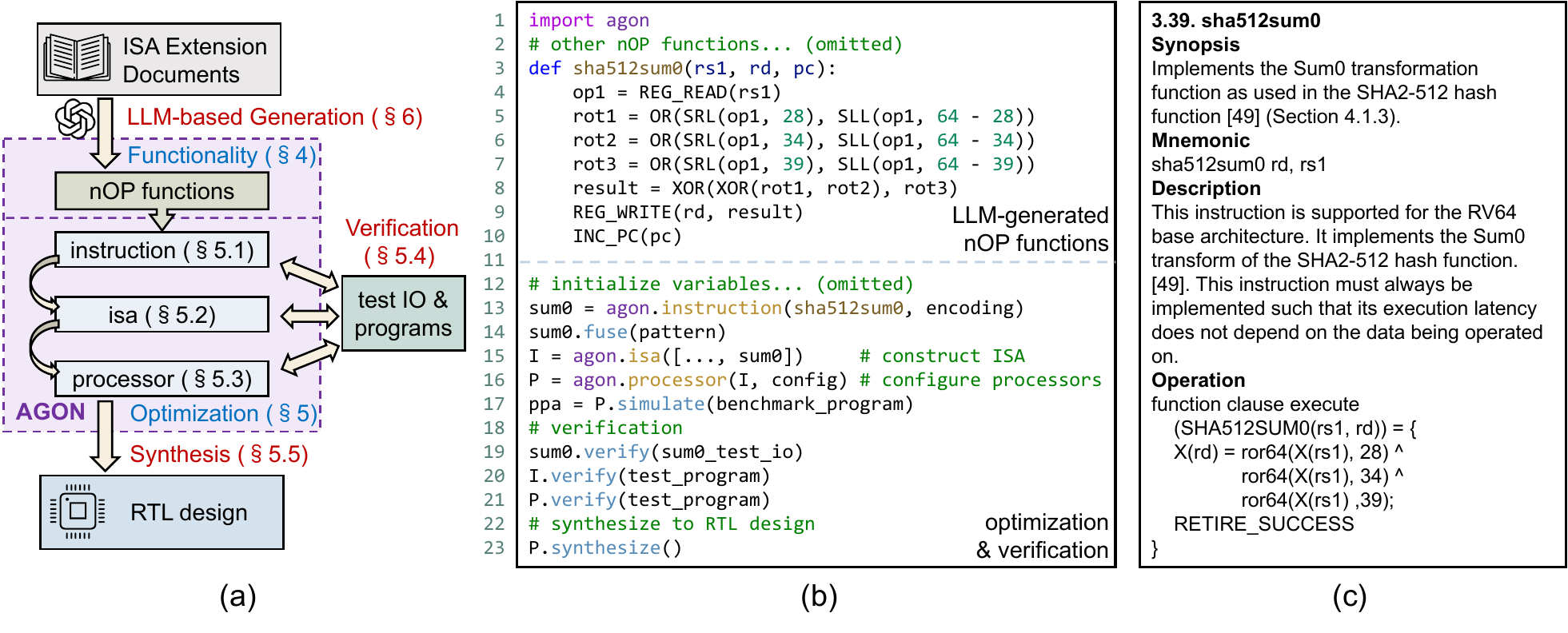}
    \caption{The overview of \xname. (a) \xname decouples optimization from functionality, using nOP functions to describe the processor functionality, and then instantiates, optimizes, and verifies designs at different levels through a set of primitives. (b) An example of using AGON for design. The functional descriptions are generated with LLM. (c) The ISA document provided to LLM for the \code{orange}{sha512sum0} instruction in (b). This snippet is taken from~\cite{riscv_crypto}, which originally contains natural language descriptions and pseudo code.}
    \label{fig:agon_overview}
\end{figure*}

\section{Background and Motivation}
\label{sec:background}

\subsection{Customized Processors}
Customized processors have been widely used in scenarios demanding highly optimized performance for specific tasks, especially with area or power constraints such as embedded systems~\cite{eusse2014flexible}, signal processing~\cite{liao2012high,liu2005application}, networking~\cite{shahabuddin2021asip,van2006scalable,de2007ultra}. 
Compared to out-of-core accelerators, customized processors offer the advantages of minimal area overhead, low data transfer overhead, and ease of programming, making them an attractive choice~\cite{li2023append}.
The key is to utilize instruction set extensions to combine the flexibility of general-purpose processors and the high performance of application-specific integrated circuits.
These extensions encompass a range of functionalities, including vector extensions~\cite{riscv_v}, security extensions~\cite{riscv_crypto, harris2021morpheus, menon2017shakti}, virtualization extensions~\cite{chen2022duvisor}.
However, supporting customized instruction extensions entails meticulous design and modification of the processor architecture.
It demands intensive coordination among architecture design, circuit development, and verification teams, which incurs significant time and human efforts, making customization costly. 

\subsection{Automated Processor Design Techniques}

High-level synthesis tools enable designers to describe circuit functionality at a high level of abstraction using languages like C, C++, or System C~\cite{zhang2021eniad,blott2018finn}.
These tools automatically generate hardware description languages from high-level specifications, accelerating the design process and enabling rapid exploration of design alternatives. 
HLS techniques first analyze the algorithm described in high-level code written by experts then partition it into hardware-friendly components, and perform data flow analysis to identify opportunities for parallel execution and optimization, such as pipelining and parallelized loops. Although HLS techniques significantly reduce development time, the PPA from HLS is typically inferior to those from traditional RTL design flow~\cite{8356004}.

In more practical usages, designers leverage Chisel~\cite{bachrach2012chisel} for agile processor developing~\cite{zhao2020sonicboom, xu2022towards, asanovic2016rocket}. 
Chisel, a hardware construction language embedded in Scala, leverage an intermediate representation called FIRRTL~\cite{izraelevitz2017reusability}, which can be optimized and synthesized into RTL. This results in hardware designs that can achieve comparable performance to those developed in Verilog. However, utilizing Chisel requires designers to possess expertise and write substantial amounts of code.
Automated processor design with Chisel still faces considerable challenges and requires further development to streamline the process.

\subsection{LLM-based Attempts on Hardware Generation}
\label{subsec:backgroud-llm-hdl-gen}

Recent researches~\cite{Blocklove_2023, chang2023chipgpt, lu2023rtllm, thakur2023benchmarking, liu2023verilogeval, chang2024data, pei2024betterv} have explored the potential of LLMs for directly generating RTL code from natural language description. These studies can be divided into two categories: how to utilize existing LLMs to assist in generating circuits without training, and how to fine-tune open-source LLMs for better module-level HDL generation. For the former, Chip-Chat~\cite{Blocklove_2023} and Chip-GPT~\cite{chang2023chipgpt} employ ChatGPT to generate circuits in a conversational manner which requires interactive human feedback. RTLLM~\cite{lu2023rtllm} utilizes a self-planning technique, where LLM first gives out a high-level plan and then generates the HDL code. However, even equipped with GPT-4, RTLLM fails to generate functionally correct HDL code for simple designs, such as a 32-bit adder. For fine-tuning LLMs, Thakur et al.~\cite{thakur2023benchmarking} and Liu et al.~\cite{liu2023verilogeval} fine-tune different LLMs with up to 16B parameters for HDL generation and presented an evaluation suite, respectively. Chang et al.~\cite{chang2024data} and BetterV~\cite{pei2024betterv} present data augmentation methods to collect high-quality Verilog datasets, and fine-tune different open-sourced models. 
None of these methods have demonstrated accuracy that fully exceeds GPT-4 and they can generate module-level HDL only, which is still far from assisting in designing chips.

There are mainly three gaps that prevent LLM from designing practical-level processors using HDLs:

\textbf{The gap between the long code processing ability of LLMs and the scale of processor design projects.}
LLMs struggle to process information within long contexts effectively, especially when their length exceeds their training data~\cite{liu2024lost}. The average length of Verilog code in the RTLCoder~\cite{liu2023rtlcoder} fine-tuning dataset is 451.9 tokens, while the main part of open-source in-order CPU core cv32e40p~\cite{gautschi2017near} comprises 28 SystemVerilog files totaling 243,013 tokens.
Evaluations in RTL-Repo~\cite{allam2024rtlrepo} show that as the input context length increases from 2k to 64k tokens, the accuracy of GPT3.5 and RTLCoder-Deepseek generating the next Verilog code line drops by 54.0\% and 50.1\%, respectively.
Also, for complex code projects, LLMs struggle with complicated cross-file dependencies~\cite{liu2024repobench, zhang-etal-2023-repocoder, allam2024rtlrepo}. 
Designing processors using HDLs necessitates exhaustive files with complex inter-module calling relationships. 
These gaps in LLM capabilities hinder the application of LLMs in processor design using HDL.

\textbf{The gap between HDLs and other software languages.}
LLMs leverage code as part of their training data to enhance their logical and coding capabilities. However, HDLs focus on describing circuit structures, which is distinct from software languages that focus on program control-data flow. Moreover, high-quality HDL codes typically belong to integrated circuit companies, making training data less.
HDLs account for only about 1/800 of all programming languages~\cite{li2023starcoder}. Consequently, LLM's generation capabilities in HDL are significantly inferior to those in software programming languages, with respective state-of-the-art methods scoring 46.1~\cite{pei2024betterv} on VerilogEval~\cite{liu2023verilogeval} compared to 96.9~\cite{zhong2024ldb} on HumanEval~\cite{chen2021evaluating}, a Python generation benchmark. HDLdebugger~\cite{yao2024hdldebugger} identifies the frequent reasons for LLM's errors in generating Verilog, most of which are related to port connections, sequential structures, and bit width management. This highlights a clear semantic gap between HDL and software languages in these aspects, where LLMs face challenges in bridging the gap.

\textbf{The gap between HDL code to hardware PPA.}
The chain from HDL code to PPA evaluation for hardware is lengthy, requiring complex and time-consuming simulation and synthesis to obtain architectural or circuit-level PPA evaluations~\cite{zhong2023llm4eda}. Accurately predicting PPA in the early stages of the EDA flow is extremely challenging~\cite{wu2022survey}. Additionally, the downstream optimization goals from HDL code to physical design are diverse, usually requiring multiple iterative cycles to refine the design in industrial workflows~\cite{pei2024betterv}. LLMs lack an understanding of the relationship between HDL code and PPA metrics due to a lack of code-metric data pairs as training data~\cite{zhong2023llm4eda}, and they also cannot foresee diverse downstream optimization goals, impeding the optimization of HDL code.

\textbf{In summary}, to achieve efficient automated processor design, we should \circleone achieve design efficiency across diverse customized applications, \circletwo guarantee functional correctness with minimal human intervention, and \circlethree empower hardware optimization to attain comparable performance with human-crafted processors. 
LLM-assisted methods unveil the glimmer of automated processor design from natural language specifications to hardware implementations, while still maintaining some fundamental gaps. 
By effectively solving the problems, \xname provides a feasible infrastructure for automated processor design prototyping in the era of LLMs.

\section{\xname Overview}
\label{sec:overview}
\xname is a comprehensive framework designed for LLM-based processor design, addressing the aforementioned issues. Figure~\ref{fig:agon_overview} illustrates the overall framework of \xname and showcases an example using \xname to design a customized processor.
AGON decouples functionality from optimization. From the functionality aspects (Section~\ref{sec: nOP Design Methodology}), it utilizes nOP functions generated by LLM to define the functionality of instructions in the processor. From the optimization aspects (Section~\ref{sec:backend}), it contains three levels of implementation: instruction (Section~\ref{subsec:instr_schedule}), ISA (Section~\ref{subsec:isa_schedule}), and processor (Section~\ref{subsec:processor_schedule}). It utilizes a set of primitives for instantiation, optimization, and verification(Section~\ref{subsec:verification}), keeping the functionality described in nOP functions unchanged, and eventually implements the processor to an RTL design (Section~\ref{subsec:hardwaregen}). By leveraging LLMs to generate nOP functions from the ISA documents (Section~\ref{sec:frontend}) and utilizing AGON's auto-tuning primitives, designers can efficiently design customized processors with extended instructions, requiring minimal human intervention.

Figure~\ref{fig:agon_overview} (b) illustrates a simplified example of using AGON to design a customized processor. \xname is embedded in Python, and the design starts with LLM-generated \opname functions (lines 2-10), which are used to define the functionalities of the instructions. The \opname function and its encoding are instantiated as a single \code{purple}{instruction} (line 13), followed by optimizations at the instruction level (lines 14). A set of instructions is then combined into an \code{purple}{isa} (line 15) and configured as a \code{purple}{processor} (line 16). We can \code{teal}{simulate} (line 17) and \code{teal}{synthesize} (line 23) the processor. We \code{teal}{verify} at multiple levels based on test IO and programs (lines 19-21).

We will then elaborate in the following sections how \xname, through its unique design of \opname functions and the decoupling of functionality and optimization, achieves \circleone expressive and efficient, \circletwo correctness assurance, and \circlethree PPA optimization capabilities.

\section{Functionality Descriptions in \xname}
\label{sec: nOP Design Methodology}

To bridge the gap between natural language-based specifications and hardware implementation, we propose a novel intermediate representation (IR) named \opname functions in \xname for processor functionality definitions.

\begin{figure}[t]
    \centering
    \includegraphics[width=1\linewidth]{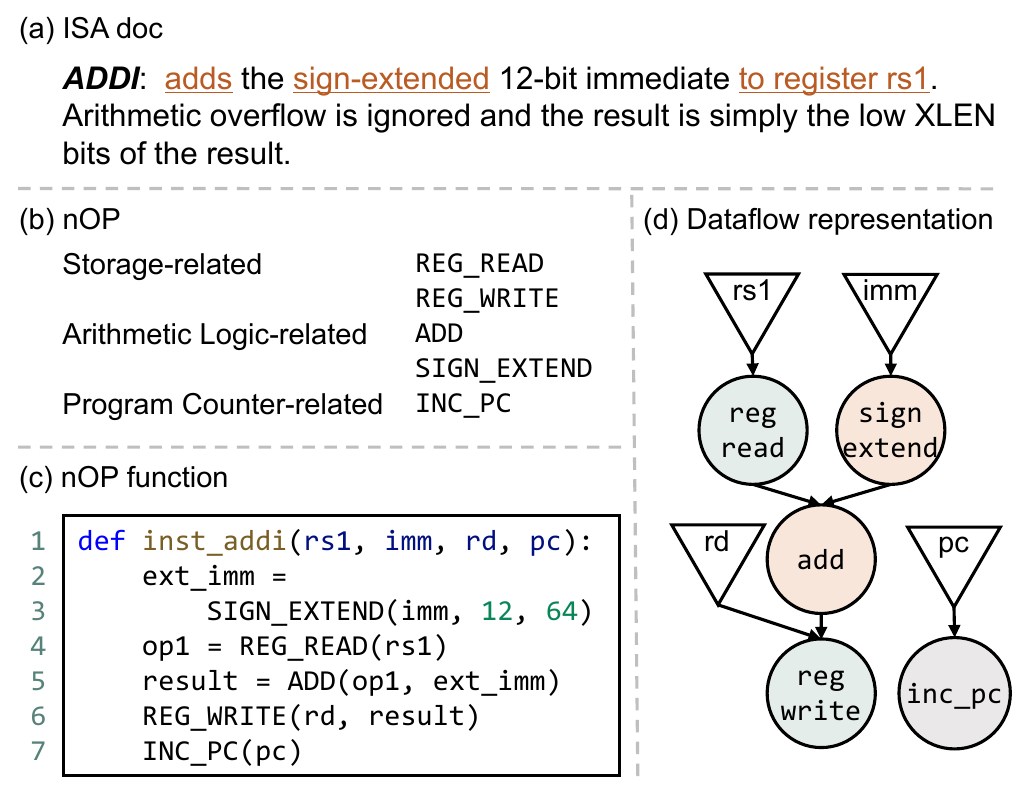}
    \caption{Concepts and examples about \opnames, \opname functions and \opname dataflow graphs. The \opnames in (b) correspond to the operations described in the natural language document in (a), and can be combined to form the \opname function in (c), and the dataflow is shown in (d).}
    \label{fig:concept_explaination}
\end{figure}

\textbf{\opname function is expressive in describing instruction functionalities.} \opnames are derived from natural language descriptions in ISA documents, and considering common hardware modules. Figure~\ref{fig:concept_explaination} (a), (b), and (c) illustrate the relationship between ISA documents, \opnames, and \opname functions, with the dataflow shown in Figure~\ref{fig:concept_explaination} (d), using the \code{orange}{addi} instruction as an example. As shown in Table~\ref{tab:nop_table}, \xname provides three categories of \opnames: storage-related, program counter (PC)-related, and arithmetic logic (AL)-related. Generally, the storage-related \opnames are used to access general registers and memory; the PC-related \opnames are used to update the PC; and the AL-related \opnames include arithmetic operations, bit operations, shift operations, comparison operations, interception and concatenation operations, and conditional assignments. For instructions in von Neumann architectures, these three types of \opname cover the overall functional space of instructions, thus enabling combining \opnames into \opname functions to define various instruction functionalities. \opnames are plug-and-play for \xname and can be easily extended by simply providing the function of the new \opname and its corresponding circuit module. \opname functions are written in a Pythonic manner, and an \opname function compiler is developed to support various syntaxes, including fix-count for-loops, nested nOP calls, and built-in operations on constants like \code{blue}{+}, \code{blue}{-}, and \code{blue}{<<}, \etc These features enhance the expressiveness of using \opname functions to describe instruction functionalities.
\textbf{\opname functions serve as an efficient IR.} When designing processors using HDL, the tight coupling of instruction functionality design and the processor architecture design results in complex HDL codes. By using \opname functions, one can focus on the functionality of individual instructions. These \opname functions are also executable and can then be combined to define the execution model of a processor at the ISA level.
\begin{table}
  \caption{\opnames in \xname and their descriptions, categorized into three types.}
  \label{tab:nop_table}
  \centering
  \resizebox{\columnwidth}{!}{%
  \begin{tabular}{p{4.0cm} p{5.6cm}}
    \hline
    \opnames&Descriptions\\
    \hline
    \hline
    \multicolumn{2}{c}{Storage Related}\\
    \hline
    \code{black}{REG\_READ}(X) & Returns the value of register X. \\
    \code{black}{REG\_WRITE}(X, Y) & Writes value Y to the register X.\\
    \code{black}{MEM\_READ}(X, Y) & Reads Y-bytes from memory address X.\\
    \code{black}{MEM\_WRITE}(X, Y, Z) & Writes Y-bytes Z to memory address X.\\
    \hline
    \multicolumn{2}{c}{Program Counter Related}\\
    \hline
    \code{black}{INC\_PC}(PC) & Increases PC by 4. \\
    \code{black}{UPDATE\_PC}(X) & Updates PC to X.\\
    \code{black}{COND\_UPDATE\_PC}(C, X, PC) & If C is 1, updates PC to X, otherwise increases PC by 4.\\
    \hline
    \multicolumn{2}{c}{Arithmetic Logic Related}\\
    \hline
    
    \code{black}{AND/OR/XOR}(X, Y) & Bitwise AND/OR/XOR of X and Y. \\
    \code{black}{NOT}(X) & Bitwise NOT of X.\\
    \code{black}{ADD/SUB}(X, Y) & Returns X$+$Y / X$-$Y. \\
    \code{black}{SIGNED\_MUL}(X, Y) & Returns $\mathrm{X}_s \times \mathrm{Y}_s$.\\
    \code{black}{SLL/SRL}(X, Y) & Logical left/right shifts X by Y bits. \\
    \code{black}{SRA}(X, Y) & Arithmetic right shifts X by Y bits.\\
    \code{black}{SLICE}(X, Y, Z) & Returns X[Y:Z]. \\
    \code{black}{CONCAT}(X, Y, Z, W) & Concatenates the Y-bits X and W-bits Z. \\
    \code{black}{SIGN\_EXTEND}(X, Y, Z) & Sign-extends the Y-bits X to Z bits. \\
    \code{black}{UNSIGN\_EXTEND}(X, Y, Z) & Zero-extends the Y-bits X to Z bits.\\
    \code{black}{CMP\_GE\_S/U}(X, Y) & If X${\geq}_{s/u}$Y returns 1, otherwise 0.\\
    \code{black}{CMP\_GT\_S/U}(X, Y) & If X${>}_{s/u}$Y returns 1, otherwise 0.\\
    \code{black}{CMP\_LT\_S/U}(X, Y) & If X${<}_{s/u}$Y returns 1, otherwise 0.\\
    \code{black}{CMP\_LE\_S/U}(X, Y) & If X${\leq}_{s/u}$Y returns 1, otherwise 0.\\
    \code{black}{CMP\_NE}(X, Y) & If X$\neq$Y returns 1, otherwise 0.\\
    \code{black}{CMP\_EQ}(X, Y) & If X$=$Y returns 1, otherwise 0.\\
    
    \code{black}{COND\_ASSIGN}(C, X, Y) & If C is 1, returns X, otherwise Y.\\
    
     \hline
\end{tabular}%
}
\end{table}

Moreover, \opname functions bridge the gap between HDL and high-level programming languages. \circleone Combining all \opnames in a data-flow manner, \opname functions efficiently avoid the need for explicit interface management between circuit modules. Module ports are automatically connected during the \xname synthesis stage, based on the data flow in the \opname function. This also eliminates the need for HDL's unique sequential features, such as the \textit{blocking assignment} and \textit{unique driver} requirements for registers and finite state machine definition, among others. \circletwo \opnames only expose bit width information if necessary, such as performing signal interception (\code{black}{SLICE}), concatenation (\code{black}{CONCAT}), and extension (\code{black}{SIGN/UNSIGN\_EXTEND}), \etc For other \opnames, input and output bit widths are inferred implicitly through a type system.

In summary, the above-mentioned features enhance the expressiveness and efficiency of \opname functions, minimize their usage overhead in design, and make them more suitable for LLM generation. As detailed in Section~\ref{sec:frontend}, LLMs can accurately generate \opname functions with minimal human intervention.

\section{Primitives in \xname}
\label{sec:backend}
\xname decouples functionality and optimization, gradually implements the nOP functions into the processor RTL design through a set of primitives (listed in Table~\ref{tab:schedule_primitive}), and performs optimization and verification during the process.
\xname applies rule-based transformations to implement and optimize the processor, ensuring that the functionality of the instructions defined in the nOP functions remains unchanged, and only determines how these instructions are executed on the processor, guaranteeing process correctness. Different combinations of primitives yield different processor implementations, enabling designers to explore the design space at multiple levels. \xname employs analytical models and cycle-accurate simulators to estimate the processor's PPA and provides PPA-aware cross-level auto-tuning methods for automatic optimization.
The following of this section will detail the implementation of the \xname primitives.

\begin{table}
\caption{\xname's scheduling primitives, including instantiation, optimization, and verification.}
\label{tab:schedule_primitive}
\centering
\resizebox{\columnwidth}{!}{%
\begin{tabular}{ll}
\hline
Primitives  & Descriptions                                     \\ \hline \hline
\code{purple}{agon}.\code{purple}{instruction}(\code{black}{n, e})   & Instantiate the nOP function \code{black}{n} with its \\&corresponding encoding format \code{black}{e}.                     \\
\code{purple}{instruction}.\code{teal}{merge}\code{black}{(c)}   & Merge redundant nOP call \code{black}{c}.                     \\
\code{purple}{instruction}.\code{teal}{fuse}\code{black}{(p)}    & Fuse nOPs into one based on pattern \code{black}{p}. \\
\code{purple}{instruction}.\code{teal}{verify}\code{black}{(t)}  & Verify the instruction based on test IO \code{black}{t}.      \\ \hline
\code{purple}{agon}.\code{purple}{isa}\code{black}{([i\_0, i\_1,...])}   & Instantiate an ISA with a list of instructions \\& \code{black}{[i\_0, i\_1,...]}.                     \\
\code{purple}{isa}.\code{teal}{extend}\code{black}{(i)}          & Extend the ISA to support instruction \code{black}{i}.        \\
\code{purple}{isa}.\code{teal}{simulate}\code{black}{(b)}        & Simulate the execution of benchmark \\&program \code{black}{b} at ISA level.  \\
\code{purple}{isa}.\code{teal}{auto\_fuse}\code{black}{(g)}        & Automatically fuse nOPs across the ISA \\&with a gain of no less than threshold \code{black}{g}.                       \\
\code{purple}{isa}.\code{teal}{verify}\code{black}{(t)}          & Verify the ISA using the test program \code{black}{t}.  \\ \hline
\code{purple}{agon}.\code{purple}{processor}(\code{black}{I, c})   & Instantiate a processor with ISA \code{black}{I} and \\&configuration \code{black}{c}.                     \\
\code{purple}{processor}.\code{teal}{config}\code{black}{(c)}    & Configure the processor with configuration \code{black}{c}.  \\
\code{purple}{processor}.\code{teal}{simulate}\code{black}{(b)}  & Simulate the execution of benchmark \\&program \code{black}{b} at processor level.  \\
\code{purple}{processor}.\code{teal}{auto\_config}\code{black}{(f)}       & Explore the design space with target function \\&\code{black}{f} and configure the processor. \\ 
\code{purple}{processor}.\code{teal}{verify}\code{black}{(t)}    & Verify the processor using the test program \code{black}{t}.  \\
\code{purple}{processor}.\code{teal}{synthesize}\code{black}{()} & Synthesize the processor to RTL design.                       \\ \hline

\end{tabular}%
}
\end{table}

\subsection{Instruction-level Primitives}\label{subsec:instr_schedule}

To perform subsequent optimizations and synthesis, we first instantiate the nOP function using \code{purple}{agon.instruction} primitive. \xname's built-in compiler constructs a data flow graph from the nOP function. It unrolls all fixed-count for-loops, flattens nested nOP calls, and pre-computes operations with constants, converting variable-related operations into nOP calls.
The instruction's encoding format is also specified. If the compiler encounters issues it cannot resolve, it will raise an error at this stage to help designers debug.
After the instantiation, with a complete data flow graph, the instructions are already executable. Repeated calls can be merged using \code{purple}{instruction}.\code{teal}{merge} primitive if specified.

nOPs are expressive and efficient when describing instruction functionality; however, complexities arise when mapping nOPs to hardware.
We focus on mapping AL-related nOPs to the processor's arithmetic logic units (ALUs).
Implementing a functional unit (FU) for each nOP type can enhance hardware reuse: given the limited types of nOPs, complex instruction functionalities can also be realized by scheduling a small range of FUs, where the number of FU types remains independent of the number of instructions. However, this approach requires allocating at least one clock cycle for each nOP and may cause timing waste. 
We synthesize all AL-related nOPs using the 65nm technology to estimate their PPA. When setting the target frequency as 2 GHz with the clock cycle being 0.5 ns, most nOPs have too much positive timing slack, with the most affected ones listed in Table~\ref{tab:nOP_area_latency}. This will result in excessive cycle delay in instruction execution.

\begin{figure}[ht]
    \centering
    \includegraphics[width=\linewidth]{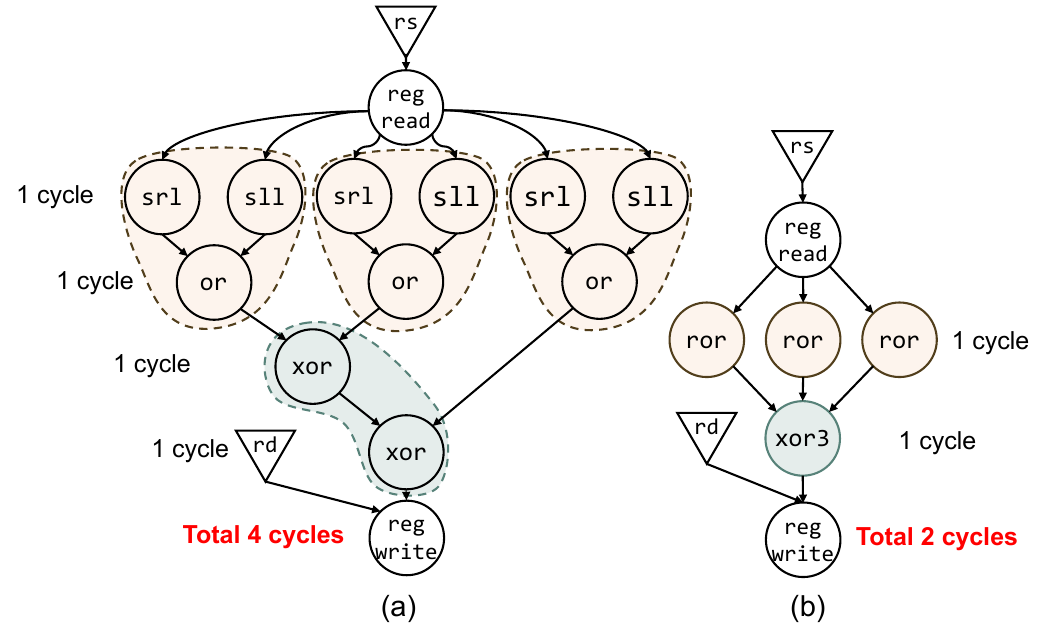}
    \caption{
    Fusing \opnames with \code{purple}{instruction}.\code{teal}{fuse} primitive. (a) The original \opname dataflow graph of instruction \code{orange}{sha512sum0}. The minimum instruction execution cycle delay is 4 cycles. (b) After fusing with two patterns \code{black}{ror} and \code{black}{xor3}, the delay is reduced to 2 cycles.}
    \label{fig:fuse_node}
\end{figure}

We design \code{purple}{instruction}.\code{teal}{fuse} primitive to fuse common nOP calling patterns into a single OP and implement it as one FU, thus the instruction execution cycle delay is shortened.
We can estimate the cycle delay optimized by only considering the total arrival time of the critical path in the pattern, as shown by Equation~\ref{eq:cycle_delay}, where $cp(g)$ stands for critical path in the pattern, $\mathcal{T}_{arr}$ stands for arrival time of the \opname, and $\mathcal{T}_{clk}$ stands for cycle time. Figure~\ref{fig:fuse_node} illustrates an example when fusing nOPs in an extended instruction \code{orange}{sha512sum0}. The cycle delay of the whole instruction before fusion is 4 cycles and is optimized to 2 cycles when applying two fusion patterns.

However, the fusion of nOPs creates dedicated FUs, potentially reducing hardware reuse and increasing the processor's area. Therefore, designers need to comprehensively consider the benefits of nOP fusion. In the next section, we will introduce a primitive that performs PPA-aware automatic pattern search and fusion at the ISA level.

\begin{table}
\caption{Synthesis results of selected AL-related \opnames.}
 \label{tab:nOP_area_latency}
 \centering
 \resizebox{\columnwidth}{!}{%
 \begin{tabular}{lccc}
 \hline
\opname & area ($um^2$) & arrival time ($ns$) & slack ($ns$) \\
\hline\hline
\code{black}{UNSIGN\_EXTEND} & 1037.5 & 0.09 & 0.41 \\
\code{black}{COND\_ASSIGN} & 1931.8 & 0.18 & 0.32 \\
\code{black}{CONCAT} & 2061.4 & 0.27 & 0.23 \\
\code{black}{AND} & 1805.8 & 0.12 & 0.38 \\
\code{black}{OR} & 1968.1 & 0.12 & 0.38 \\
\code{black}{NOT} & 1257.5 & 0.09 & 0.41 \\
\code{black}{XOR} & 2309.0 & 0.16 & 0.34\\
\hline
\end{tabular}%
}
\end{table}

\begin{align}
\mathcal{D}_{cycle}(g) & = \left \lceil \frac{\sum_{o\in cp(g)}\mathcal{T}_{arr}(o)}{\mathcal{T}_{clk} }  \right \rceil   \label{eq:cycle_delay}
\end{align}

\subsection{ISA-level Primitives}\label{subsec:isa_schedule}

We use a set of instructions to instantiate an \code{purple}{agon.isa}, and we can additionally extend support for new instructions using \code{purple}{isa}.\code{teal}{extend} primitive.

The difference between an \code{purple}{agon.isa} and a simple collection of \code{purple}{agon}.\code{purple}{instruction}s lies in its capability to simulate program execution at this level. The \code{purple}{isa}.\code{teal}{simulate} primitive simulates executing one instruction at a time, determining the next instruction to execute based on the PC value after execution. This facilitates analyzing the execution results and reporting the dynamic instruction counts of a program.

As mentioned above, fusing nOPs requires appropriate patterns, while identifying all these patterns is challenging. We propose an automatic pattern searching and fusion primitive at the ISA level. We develop a greedy PPA-aware pattern search strategy to collect patterns across the ISA, outlined in Algorithm~\ref{alg:find_template}. We traverse all subgraphs (reflecting local nOP calling relationships) in the \opname data flow graph across all instructions supported by the ISA, estimating the cycle delay after fusing with Equation~\ref{eq:cycle_delay}, and identify subgraphs that bring at least $gainThreshold$ timing optimizations. We merge isomorphic subgraphs into a single pattern and then use a greedy approach to apply these patterns according to their sizes to all instructions that are supported by the ISA.
\begin{algorithm}
\caption{Collecting patterns when using the \code{purple}{isa}.\code{teal}{auto\_fuse} primitive.}\label{alg:find_template}
\begin{algorithmic}[1] 
\renewcommand{\algorithmicrequire}{\textbf{Input:}}
\renewcommand{\algorithmicensure}{\textbf{Output:}}
\Require all {\opname}Graphs, gain threshold gainThreshold
\Ensure patternSet
\State subgraphs $\gets$ \textsc{subgraphTraverse}({\opname}Graphs)
\State patternSet = ()
\For{subg \textit{in} subgraphs}
    \State \opnames =  \textsc{criticalPath}(subg)
    \State {\opname}Lats = \{ \textsc{Latency}(\opname) \textbf{for} \opname \textit{in} \opnames \} 
    \State {{\opname}Cycles} = \{ \textsc{Ceil}({{\opname}Lat / cycleTime}) \\
                            \qquad  \qquad \qquad \qquad \qquad \qquad  \textbf{for} {{\opname}Lat} \textit{in} {{\opname}Lats} \}
    \State {initCycle} = \textsc{Sum}({{\opname}Cycles})
    \State {optCycle} = \textsc{Ceil}(\textsc{Sum}({{\opname}Lats}) / {cycleTime})
    
    \State gain $\gets$ 1 - optCycle / initCycle
    \If{gain $\geq$ gainThreshold}
        \State \text{patternSet.\textit{add}}(subg)
    \EndIf
\EndFor
\State \Return patternSet
\end{algorithmic}
\end{algorithm}

\subsection{Processor-level Primitives}\label{subsec:processor_schedule}
AGON instantiates an out-of-order(OoO) processor based on the ISA and configures it with a set of hardware parameters (as shown in Table~\ref{tab:design_space_para}). We take the nOPs within the instructions as the basic dynamic scheduling unit for the OoO processor. The processor dynamically executes the nOPs based on the nOP dependencies. Compared to scheduling based on instructions, this implementation better reuses FUs. AGON supports a comprehensive OoO processor microarchitecture design at the processor level, including a decoupled frontend, a branch prediction mechanism, and a memory hierarchy including multi-level caches. AGON provides a cycle-accurate OoO processor simulator that simulates the execution of benchmark programs, reports dynamic cycle counts, and estimates the area and power of the processor using McPAT. Designers can explore different optimized configurations based on the PPA reported by the simulator.

However, Table~\ref{tab:design_space_para} defines a space containing over 179 million design points, making manual exploration difficult. We further provide a \code{purple}{processor}.\code{teal}{auto\_config} primitive for automatic optimization of the configurations. This design space exploration (DSE) process can be formally represented as Equation \ref{eq:dse}, where $x$ denotes a design point in the design space, represented by an array of hardware parameters, and $f$ signifies the cost function. To facilitate efficient exploration, AGON employs a parallelized Bayesian optimization algorithm for searching.
\begin{align}
x^*&=\mathop{\arg\min}\limits _{x}f(cycle(x), area(x), power(x))\label{eq:dse}
\end{align}

\begin{table}
  \caption{The design space for \code{purple}{processor}.\code{teal}{auto\_config} primitive.}
  \label{tab:design_space_para}
  \centering
  \resizebox{0.8\columnwidth}{!}{%
  \begin{tabular}{ccccl}
    \hline
    Module&Parameter&Candidate value\\
    \hline
    \hline
    Branch & BTB size & [128, 256, 512] B\\
    Prediction & replace policy & [lfu, fifo, random, lru]\\
    \hline
    Fetch & fetch width & [4, 6, 8]\\
     & fetch buffer entry&[16, 32, 48]\\
    \hline
    Decode & decoder width & [1, 2, 3, 4, 5]\\
    \hline
    Execute & ROB entry& [32, 64, 128, 256]\\
     & load queue entry & [8, 16, 32]\\
     & store queue entry& [8, 16, 32]\\
     & BJU number & [1, 2]\\
     & AGU number & [1, 2]\\
     & ALU number & [1, 2, 3, 4]\\
     & EXT number & [1, 2, 3, 4]\\
    \hline
    Cache & cache way & [4, 8]\\
     & block size & [32, 64] B\\
     & L1D size & [8, 16, 32] KB\\
     & L2D size & [128, 256, 512] KB\\
     & L3D size & [1, 2, 4, 8] MB\\
     \hline
\end{tabular}%
}
\end{table}

\subsection{Multi-level Verification}\label{subsec:verification}
Verification is crucial to the processor design. In \xname, we offer multi-level verification mechanisms from instruction to ISA model and down to architectural levels, utilizing test IOs or test programs to verify the design.

As mentioned above, after instantiating the nOP function into an \code{purple}{agon}.\code{purple}{instruction}, it becomes executable. Designers use \code{purple}{instruction}.\code{teal}{verify} to validate the correctness of the nOP function by providing standard inputs and outputs and comparing them with the execution results of the instruction. It is worth noting that since the instructions involve memory-related operations, we use initial states that can simulate register and memory access traces as inputs, and use the transition states after instruction execution as outputs, rather than merely passing values.

When instructions are instantiated to an \code{purple}{agon}.\code{purple}{isa}, \xname provides ISA-level simulation of a program, thereby also supporting verifying the design with a test program. 
Verification at the ISA model level does not involve the processor's microarchitecture; it only verifies the functionality of the instructions in the ISA, without involving architectural details, such as branch prediction, memory hierarchy,~\etc

After the \code{purple}{agon}.\code{purple}{processor} is instantiated, the microarchitecture is introduced via configuration. 
We utilize \xname's built-in simulator to fully simulate the execution of the test programs on the processor, and verify the correctness at the processor level, including ISA definitions and microarchitecture implementations.

\subsection{Synthesize to RTL}\label{subsec:hardwaregen}

After the optimization and verification, we synthesize the design to RTL implementations on top of the BOOM project~\cite{zhao2020sonicboom}. We integrate \opname FUs into multifunctional units to reduce register ports. For ease of implementation, we temporarily only support fusing all AL-related \opnames within each instruction when synthesis. The fused OPs are then automatically converted to Chisel functions and integrated with BOOM. While instructions with more than two inputs execute coherently in \xname simulator, implementing them requires significant modifications to BOOM. We defer the hardware implementation of other fusing strategies and instructions with three or more inputs for future work.

\section{LLM-based \xname Generation}
\label{sec:frontend}
\xname, through its functionality optimization decoupling and \opname design, is suitable for LLMs in generating instruction descriptions. 
Nonetheless, since the training set of LLMs contains no \xname-related data, we develop a framework that employs few-shot learning, self-debugging, and sampling-selection techniques to generate accurate \opname functions while minimizing the need for human effort.

\begin{figure}[ht]
    \centering
    \includegraphics[width=0.9\linewidth]{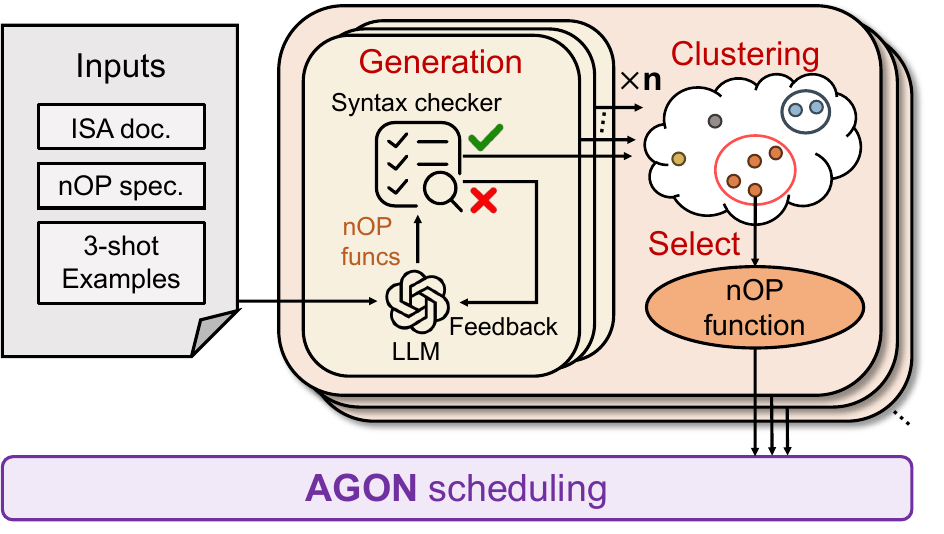}
    \caption{The \xname's \opname function generation method. After the \opname functions are generated, they are sent for \xname's scheduling.}
    \label{fig:llm_generation}
\end{figure}

Figure~\ref{fig:llm_generation} illustrates our framework for generating the \opname functions based on the natural language ISA documents.
The starting point of this framework is the natural language specifications of the instructions.
These specifications are taken directly from real-world ISA documents~\cite{riscv_crypto,riscv_p}. These informal specifications have various formats and styles. 
Some only describe the purpose of the instruction, while others provide detailed information about the operations. Some may also use pseudocode to aid in explaining functionality.

\subsection{Few-shot Learning} 
\begin{figure}[ht]
    \centering
    \includegraphics[width=\linewidth]{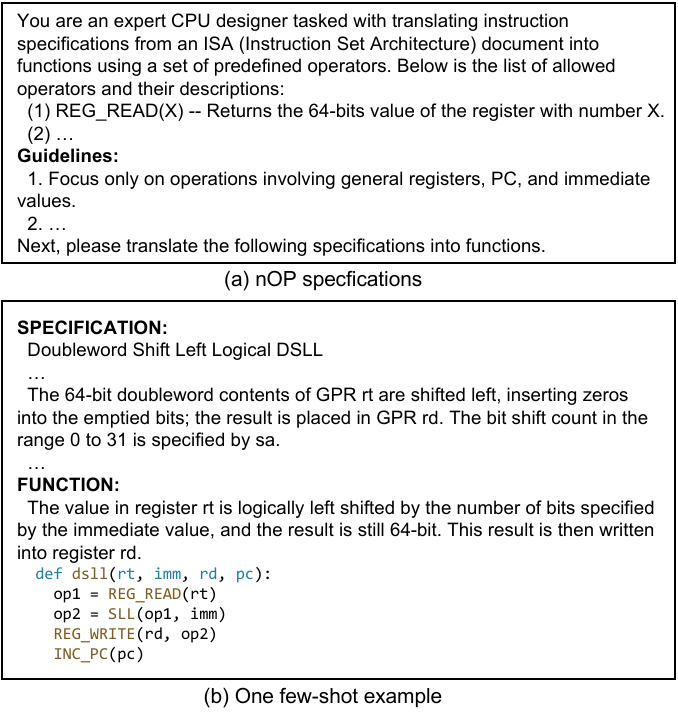}
    \caption{Illustration of \opname specifications and few-shot examples. \opname specifications are provided to LLM along with 3-shot examples to conduct few-shot learning.}
    \label{fig:nOP_generation_demo}
\end{figure}

We first provide the \opname specifications (Figure~\ref{fig:nOP_generation_demo} (a)) and \opname function regulations as prompts to the LLM. We then require the LLM to generate the corresponding \opname function according to the instruction specification, following the provided regulations.
Brown \etal~\cite{brown2020language} demonstrate that LLMs are capable of few-shot learning using examples provided in the prompt. Therefore, we provide the natural language specifications of three instructions from the MIPS instruction set, along with their corresponding hand-crafted \opname functions, as examples for LLMs to perform few-shot learning, one of which is presented in Figure~\ref{fig:nOP_generation_demo} (b). We select three MIPS instructions as examples to prevent overlap with RISC-V instructions in the target generation task, thereby ensuring a fair evaluation of our method's generalization abilities.

\subsection{Self-debugging}  
We utilize the compiler provided in \xname to check the syntax of \opname functions generated by LLMs and provide feedback when syntax errors occur~\cite{chen2023teaching}. If the \opname function fails the syntax check, the erroneous \opname function and the syntax check feedback are provided back to the LLM for regeneration. Repeat the above process until the generated \opname function passes the syntax check or the iteration times threshold is reached. 

\subsection{Sampling and Selection} 
The output of LLMs is stochastic. To enhance the quality of the \opname functions generated by the LLM, we apply a sampling-clustering-selection mechanism proposed by Li \etal~\cite{li2022competition}. We repeat the aforementioned generation and self-debugging processes $n$ times to obtain $n$ \opname functions.
Subsequently, due to the \opname functions executable feature, we generate multiple random input states, execute each \opname function with all initial states and obtain the transfer states as the outputs. 
We cluster \opname functions based on their output states. \opname functions that have the same output states are grouped into the same cluster. Eventually, an \opname function from the largest cluster is selected as the final output. We provide this clustering method instead of simply filtering \opname functions using \code{purple}{instruction}.\code{teal}{verify} primitive as comprehensive testing IOs may not always be available.

Finally, if the \opname function generated by the above framework still contains errors, we correct it manually. 
While no method can guarantee the correctness of LLM generation, our approach avoids modifying complex HDL codes.
This significantly reduces the need for expert involvement compared to other LLM-assisted circuit generation methods. The instruction set \opname functions generated through the above steps are then sent to \xname for scheduling.

\section{Experiments}
\label{sec:exp}

\begin{table}
  \caption{\xname's targeting benchmarks.}
  \label{tab:n_instr}
  \centering
  \resizebox{0.8\columnwidth}{!}{%
  \begin{tabular}{ccc}
    \hline
    &name & extended instructions\\
    \hline
    \hline
    k0 &SHA256  & \code{black}{sha256\{sig0, sig1, sum0, sum1\}}\\
    k1 &SHA512  & \code{black}{sha512\{sig0, sig1, sum0, sum1\}}\\
    p0& Fast Math & \code{black}{clz32}\\
    p1& Complex Math & \code{black}{mulsr64, smtt32, smbt32,}\\ 
    &&\code{black}{smdrs32, kmxda32, pkbb32, }\\
    &&\code{black}{kwmmul, kstsa32, ksubw, clz32}\\
    p2& Matrix Mult  & \code{black}{smmul,	smar64}\\
    p3& Matrix Add & \code{black}{kadd32,	kaddw} \\
    p4& Matrix Sub  & \code{black}{ksub32,	ksubw} \\
    p5& Matrix Scale & \code{black}{pkbb32,	kslra32,	smmul} \\
    \hline
\end{tabular}%
}
\end{table}
\subsection{Experimental Setup}

We conduct detailed experiments to evaluate \xname. We select RV64I as the base ISA and eight domain-specific applications as benchmarks, two of which are from the test suite of the RISC-V Zknh extension~\cite{riscv_crypto}, while the remaining six are from Nuclei-Software~\cite{NMSIS}, which is a business software based on the RISC-V P extension~\cite{riscv_p}. The SHA256 and SHA512 benchmarks are mostly adopted hash functions in cryptography. Fast Math benchmark provides a set of fast approximation functions to sine, cosine, and square root, and the Complex Math benchmark performs operations on complex data vectors. The last four benchmarks are designed for matrix calculations.
The involved extended instructions for these benchmarks can be found in Table \ref{tab:n_instr}.

We utilize GPT-4 as the embedded LLM in our generation method (Section~\ref{sec:frontend}) with 3-shot examples. The temperature for generation is set to 0.3. In each generation process, we allow up to 3 rounds of syntax check feedback and generate up to 9 samples for clustering. We evaluate the accuracy of AGON generation using the pass@1~\cite{chen2021evaluating} metric, as shown in Equation~\ref{eq:pass_at_k}, where $n$ stands for total generation times and $c$ stands for the number of correct generations.
\begin{align}
pass @ 1:=\mathbb{E}_{\text {problems }}\left[\frac{c}{n}\right]
\label{eq:pass_at_k}
\end{align}
We calculate the number of lines of code (LoC) required in modifying nOP functions for all instructions in a processor to be correctly implemented, to evaluate the minimum human intervention needed when using AGON.

We utilize AGON's auto-tuning primitives (\code{purple}{isa}.\code{teal}{auto\_fuse} and \code{purple}{processor}.\code{teal}{auto\_config} ) in our experiments to demonstrate AGON's optimization capabilities and its ability to minimize human effort in processor design. The dynamic execution statistics are reported by simulating the execution of the benchmarks in the \xname embedded simulator. Based on the optimal architecture obtained from \xname's auto-tuning, we implement our processor and synthesize the generated circuit to report its accurate area. All syntheses are done using a commercial EDA tool with a 65nm technology.

\begin{table}
  \caption{Pass rate comparison for LLM-based generation (in 3 trials).}
  \label{tab:agon_vs_chisel}
  \centering
  \resizebox{1\columnwidth}{!}{%
  \begin{tabular}{cccccc}
    \hline
    instruction & \xname& Chisel&instruction& \xname& Chisel\\
     & pass@1(\%)& pass@1(\%)&& pass@1(\%)& pass@1(\%)\\
    \hline
    \hline
    \multicolumn{6}{c}{RV64I}  \\
    \hline
    49 instructions & 95.9 & -\\
    \hline
    \multicolumn{6}{c}{Zknh extension} \\
    \hline
    \code{black}{sha256sig0} & \textbf{100} & 0 & \code{black}{sha256sig1} & \textbf{100} & 0 \\
    \code{black}{sha256sum0} & \textbf{100} & 0 & \code{black}{sha256sum1} & \textbf{100} & 0 \\
    \code{black}{sha512sig0} & \textbf{67.7} & 0 & \code{black}{sha512sig1} & 0 & 0 \\
    \code{black}{sha512sum0} & \textbf{100} & 0 & \code{black}{sha512sum1} & \textbf{100} & 0 \\
    \hline
    \multicolumn{6}{c}{P extension} \\
    \hline
    \code{black}{kadd32} & \textbf{100} & 0 & \code{black}{kaddw} & \textbf{100} & 67.7 \\
    \code{black}{kslra32} & 0 & 0 & \code{black}{ksub32} & \textbf{100} & 0 \\
    \code{black}{ksubw} & \textbf{100} & 0 & \code{black}{pkbb32} & \textbf{100} & 67.7 \\
    \code{black}{smar64} & \textbf{100} & 0 & \code{black}{smmul} & \textbf{100} & 0 \\
    \code{black}{clz32} & 0 & 0 & \code{black}{mulsr64} & \textbf{100} & 67.7 \\
    \code{black}{smtt32} & \textbf{100} & 0 & \code{black}{smbt32} & \textbf{100} & 0 \\
    \code{black}{smdrs32} & \textbf{100} & 0 & \code{black}{kmxda32} & \textbf{67.7} & 0 \\
    \code{black}{kwmmul} & 0 & 0 & \code{black}{kstsa32} & \textbf{100} & 0\\
    \hline
    
\end{tabular}%
}
\end{table}
\begin{table}
  \caption{Processor complexity and development effort compared with other LLM-assisted processor design methods.}
  \label{tab:dev_effort_vs_gpt}
  \centering
  \resizebox{0.9\columnwidth}{!}{%
  \begin{tabular}{ccccc}
    \hline
 design & area ($\mu m^2$) & \#instructions &debug LoC\\
    \hline
    \hline
    tiny-scale CPUs\\
    \hline
    Chip-Chat~\cite{Blocklove_2023}  & -  & 24 & 1896  \\
    ChipGPT~\cite{chang2023chipgpt}  & 3240.8  & 10  & 9  \\
    \hline
    our customized processors\\
    \hline
    SHA256 & 9.63e6 & 57 & 2\\
    SHA512 & 6.47e6 & 57 & 4.67\\
    Fast Math & 6.42e6 & 53 & 4\\
    Complex Math & 9.60e6 & 62 & 9.67\\
    Matrix Mult & 11.61e6 & 54 & 2\\
    Matrix Add & 9.96e6 & 54 & 2\\
    Matrix Sub & 9.59e6 & 54 & 2\\
    Matrix Scale & 8.54e6 & 55 & 7.33\\
    \hline
\end{tabular}%
}
\end{table}

\subsection{\xname Provides Efficiency}
To illustrate the efficiency of using AGON for processor design, we evaluate the pass rate of LLM generating AGON for describing each extended instruction, and compare it with generating corresponding dedicated ALUs in Chisel, both with 3-shot examples and providing the same ISA documents as input. 
We also report an average pass rate for 49 RV64I standard instructions. Since RV64I instructions involve memory access or branch operations, their functionality cannot be described using a single ALU; therefore, comparing them with Chisel's pass rate is not feasible.

Table~\ref{tab:agon_vs_chisel} illustrates the results.
The experimental results show that for the RV64I instructions, our average pass rate is as high as 0.96. Analyzing the generation of RV64I instructions, we find that the LLM fails only when generating the \code{orange}{lui} and \code{orange}{auipc} instructions, and successes in generating all other instructions. This proves that with the \opname function acting as an efficient IR, \xname can correctly generate almost all basic instructions without human intervention. For complicated extended instructions, \xname succeeds in generating 18 instructions without human intervention. The pass rate of \opname functions also significantly surpasses that of Chisel ALU codes in all instructions except sha512sig1, where both methods fail to generate without human modification.

We further evaluate the human effort required to modify erroneous \opname functions. 
The result is reported in Table~\ref{tab:dev_effort_vs_gpt}.
We compare our results with two other works that utilize LLMs to generate CPUs. We annotate the area and the number of supported instructions for each generated processor to illustrate their complexity. For our design, the average debugging line-of-code across 3 trials is reported to show development effort. Chip-Chat~\cite{Blocklove_2023} designs a CPU that supports 24 instructions through conversations with ChatGPT. The CPU area is not reported. In order to design such a tiny-scale CPU, they require the user to interactively provide 1896 lines of natural language prompts, which results in significant human overhead. In ChipGPT~\cite{chang2023chipgpt}, after modifying 9 lines of the generated Verilog code, a CPU that only supports 10 instructions is designed, and unable to run real-world applications. Compared with their work, equipped with \xname, LLMs are able to design processors that have significant advantages in complexity and the number of supported instructions, with lower or equivalent human interventions, and can be evaluated on real-world benchmarks, demonstrating \xname's efficiency when designing complex processors.

\subsection{\xname Enables Optimization}

\begin{figure}[ht]
  \centering
  \includegraphics[width=\linewidth]{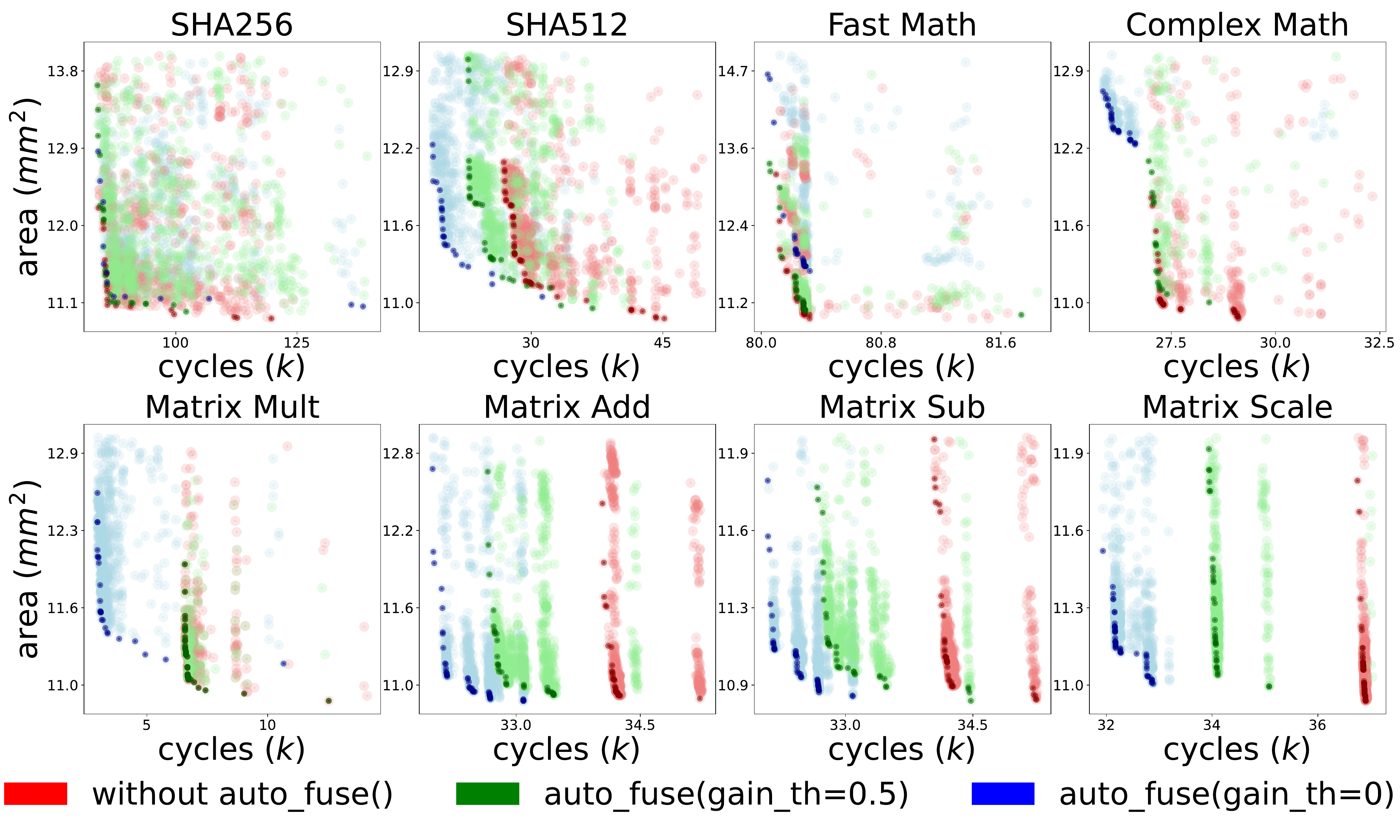}
  \caption{The results of the auto-tuning for our 8 benchmark applications. Different colors represent different settings in \code{purple}{isa}.\code{teal}{auto\_fuse} primitive, leading to different preferences in the Pareto front.
  }
  \label{fig:dse_result}
\end{figure}

Figure~\ref{fig:dse_result} shows the optimization provided by our auto-tuning primitives (\code{purple}{isa}.\code{teal}{auto\_fuse} and \code{purple}{processor}.\code{teal}{auto\_config}). We use area efficiency as the optimization goal, minimizing Equation~\ref{eq:cost_function}.
The dots represent designs that have been explored by AGON's built-in DSE engine, while the darker dots indicate that they are on the Pareto front. As analyzed previously, the choice of different nOP fusing strategies will affect the performance and area of the processors. Performing \code{purple}{isa}.\code{teal}{auto\_fuse} optimization with no gain threshold requirements (i.e. all nOPs are fused into one OP) makes it easier to obtain better performance by reducing cycle delays and register pressure, but it also leads to a larger area due to fewer hardware reuse opportunities, so the blue points tend to be distributed in the upper left of the scope. 
Applying no \code{purple}{isa}.\code{teal}{auto\_fuse} design does not require excessive dedicated computational components, so it can achieve a smaller area. However, problems of long cycle latency and register pressure will lead to lower performance, thus the red points tend to be distributed in the lower right part of the chart.
Specifying a gain threshold with a 50\% cycle reduction is a compromise, so the green points tend to be distributed in the middle. These observations are prominent in SHA512, Complex Math, and Matrix Scale benchmarks.

\begin{align}
f :&= cycle(x) \times area(x)\label{eq:cost_function}
\end{align}

In summary, the results prove that the scheduling primitives in AGON provide a large optimization space, and with auto-tuning primitives, the designers can effectively explore the design space with minimal human effort.

\begin{figure}[ht]
  \centering
  \includegraphics[width=\linewidth]{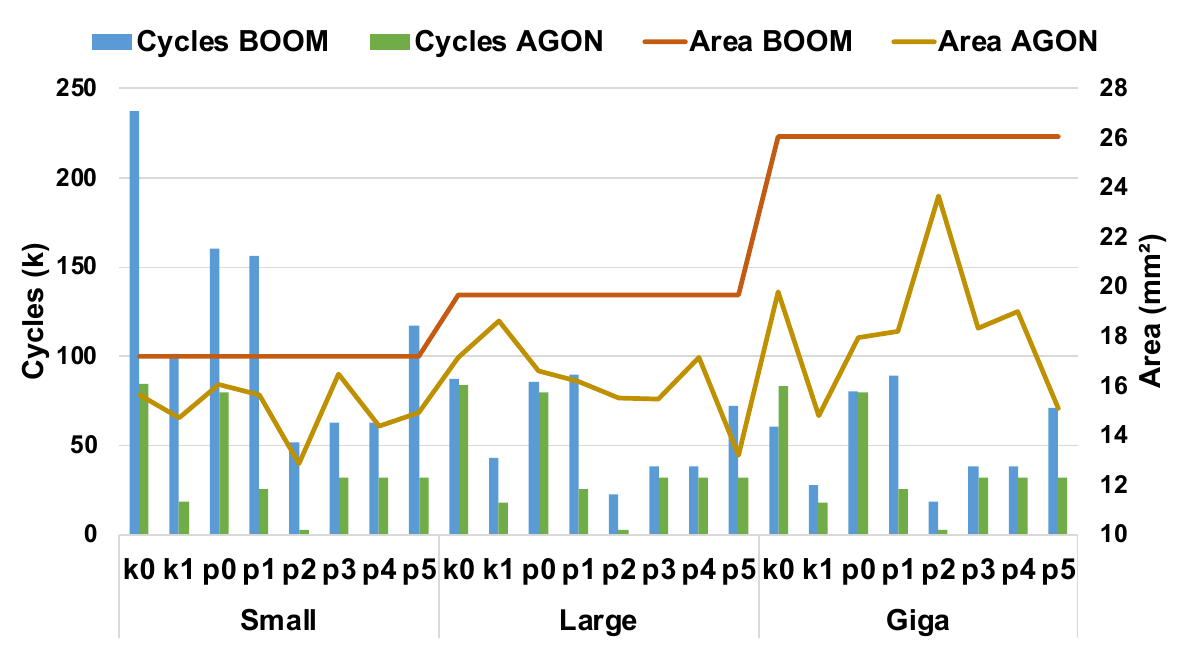}
  \caption{The cycles and area of customized processors designed with \xname and BOOM under three BOOM configurations: small, large, and giga.}
  \label{fig:vs_boom}
\end{figure}

We further compare the performance of customized processors designed using \xname with BOOM in area-constrained scenarios. To enable a fair comparison, we take the area of BOOM estimated by McPAT as the limitation, utilizing \code{purple}{isa}.\code{teal}{auto\_fuse}\code{black}{(gain\_th=0)} and \code{purple}{processor}.\code{teal}{auto\_config} primitives to optimize the performance of ASIPs with a smaller area than BOOM. Applying BOOM configurations, we use our simulator to report the performance of benchmark applications compiled to RV64I running on BOOM as the baseline. Figure \ref{fig:vs_boom} reports the cycles and area of optimized ASIPs and BOOM under three BOOM configurations: small, large, and giga. As illustrated, our approach surpasses BOOM's performance in 23 out of 24 scenarios. The average performance across all benchmarks is 2.35$\times$ faster than BOOM, while the average area is only 79.5\% that of BOOM. In scenarios with the smallest area budget (use small-BOOM's area as the limitation), we achieve an average performance improvement of 3.84$\times$ compared to BOOM and a maximum improvement of 17.40$\times$ in the processor customized for Matrix Multiplication Benchmark (p2).
The experimental results demonstrate that utilizing AGON achieves an end-to-end design and optimization process from ISA documents to processors with minimal human effort.

\section{Related Work}
\label{sec:relatedwork}

\subsection{Architecture Description Language}
Architectural description language (ADL) is an abstraction of hardware description language (HDL) in processor design. Based on the level of abstraction, it can be divided into two categories.
1) ADLs bounding to microarchitecture, including nML~\cite{van2008nml}, Lisa~\cite{schliebusch2002architecture},~\etc;
2) ADLs focusing on the description of the instruction set, including ViDL~\cite{dreesen2012vidl}, ASSIST~\cite{8806953},~\etc
Due to the absence of microarchitecture specifications in the description, it brings challenges to hardware mapping. 
Further, both categories require manual coding from designers, preventing the realization of fully automated design processes.
To the best of our knowledge, no previous work has synthesized complex out-of-order processors from ADLs.

\subsection{Assisting EDA Processes with LLM}
In addition to the aforementioned LLM-based circuit generation works, some researchers explored employing LLM to assist the EDA process~\cite{liu2023chipnemo,wu2024chateda,kande2023llm}. 
For example, ChipNemo~\cite{liu2023chipnemo} is an engineering assistant chatbot that can generate EDA scripts or analyze bugs. 
ChatEDA~\cite{wu2024chateda} generates code for manipulating EDA tools based on natural language instructions. 
Kande et al.~\cite{kande2023llm} utilize LLM to generate security assertions to assist in hardware verification.
These works do not involve the logical design of the processors and are therefore orthogonal to our work.

\section{Conclusion}
\label{sec:conclusion}
We propose \xname, a framework for LLM-based customized processor design. 
\xname decouples optimization from functionality, utilizes LLM-generated nOP functions to define the processor functionality, and provides a series of primitives to implement, optimize, and verify the design.
We evaluate the effectiveness of \xname by automatically designing a series of Out-of-Order processors, customized for different applications, with minimal human intervention.
The designed processors reach an average of 2.35 $\times$ performance improvement compared to the baseline.

AGON aims to act as an infrastructure for processor design in the era of LLMs. 
In the future, we plan to support: \circleone Other parallelized architectures, such as single-instruction-multi-data (SIMD) and multi-core processing.
\circletwo More comprehensive synthesis flow to support diverse target architectures. \circlethree More primitives to enhance optimization.
\circlefour Open-sourced AGON generation benchmarks to facilitate the research in the community.

\bibliography{references}

\vfill

\end{document}